\documentstyle[sprocl,epsfig]{article}

\newcommand \beq{\begin{eqnarray}}
\newcommand \eeq{\end{eqnarray}}
\newcommand{\be}{\begin{eqnarray}}
\newcommand{\ee}{\end{eqnarray}}

\def\del{\partial}                              

\def\frac#1#2{{#1 \over #2}}
\def\simge{\mathrel{%
   \rlap{\raise 0.511ex \hbox{$>$}}{\lower 0.511ex \hbox{$\sim$}}}}
\def\simle{\mathrel{
   \rlap{\raise 0.511ex \hbox{$<$}}{\lower 0.511ex \hbox{$\sim$}}}}

\begin{document}
\title{THE COLORED GLASS CONDENSATE AND EXTREME QCD\footnote{Talk
presented at ``Strong and Electroweak Matter 2000'', Marseille,
June 14--17, 2000.}
}

\author{E. IANCU\footnote{Address after October 1st, 2000:
Service de Physique Th\'eorique, CE Saclay,
        F-91191 Gif-sur-Yvette, France. E-mail:
eiancu@cea.fr}}

\address{CERN, Theory Division, CH-1211 Geneva 23, Switzerland}

  \maketitle\abstracts{ 
The high energy limit of QCD is controlled 
by the small-$x$ part of a hadron
wavefunction.  I argue that this part is universal to
all hadrons and is composed of a new form of matter: a Colored Glass
Condensate.  This matter is weakly interacting at very small $x$, but is
non-perturbative because of the highly occupied boson states which compose
the condensate.  Such a matter might be studied in high energy lepton-hadron
or hadron-hadron interactions.
}

\section{Introduction: Parton saturation at small-$x$}

There has been much activity in the last few years in an attempt to
understand the physics of nuclear and hadronic processes in the
regime where Bjorken's $x$ becomes very small $^{1-13}$ 
 (and References therein). A remarkable feature about this
regime is that the gluon density in the hadron wavefunction is 
so high that perturbation theory breaks down even for a small
coupling constant, because of the strong non-linear effects. 
This is quite similar to high temperature QCD
where the physics at the scale $g^2T$ is non-perturbative
because of the large thermal occupation numbers of the soft
gluons \cite{SEWM98} (and Refs. therein).

The important new  phenomenon which is expected in 
these conditions is {\em parton saturation}, which is
the fact that the density of partons per unit phase space 
$(dN/d^2p_\perp\,d\tau\,d^2x_\perp)$
cannot grow forever as $x$ becomes arbitrarilly small
(here, $\tau \equiv \ln(1/x)$). If this
growth were to occur, then the cross section for deep
inelastic scattering at fixed $Q^2$ would grow unacceptably large and
eventually violate unitarity bounds. One rather expects the cross
section to approach some asymptotic value at high energy, that is, 
to saturate.  It is also intuitively
obvious that this should happen, since if the phase space density
becomes too large, repulsive interactions are generated among the
gluons, and eventually it will be
energetically unfavorable to increase the density anyfurther. 

Saturation is expected at a phase-space density of order 
$1/\alpha_s\/$, a value typical of condensates \cite{JKMW97,KM98}.
This, together with the fact that gluons are massless bosons, 
leads naturally to the expectation that the saturated gluons
form a new form of matter, which is a Bose condensate.
Since gluons carry color which is a 
gauge-dependent quantity, any gauge-invariant formulation
will neccessarily involve an average over all colors, to restore
the invariance. This averaging procedure bears a formal resemblance to 
the averaging over background fields done for spin glasses \cite{Parisi}.
The new matter is therefore called the Colored Glass Condensate \cite{CGC}
(CGC).

In what follows, I will briefly explain the assumptions which
justify this physical picture, and the mathematical formalism which is
used to describe it.

\section{The Colored Glass Condensate}

The CGC picture holds, strictly speaking, 
only in the infinite-momentum frame, 
where the hadron propagates almost at the speed of light, and,
by Lorentz contraction, appears to the external probe
as an infinitesimally thin two-dimensional sheet located at $z=t$,
or $x^-=0$. In what follows, I shall use
light-cone (LC) vector notations: for an arbitrary
4-vector $v^\mu$, I write $v^\mu=(v^+,v^-,{\bf v}_\perp)$, with
$v^+\equiv (1/\sqrt 2)(v^0+v^3)$,
$v^-\equiv (1/\sqrt 2)(v^0-v^3)$, and ${\bf v}_\perp
\equiv (v^1,v^2)$. The dot product reads then:
$p\cdot x=p^+x^- + p^-x^+-p_\perp\cdot x_\perp\/$;
$p^-$ and $p^+$ are, respectively, the LC energy and
longitudinal momentum; correspondingly, $x^+$ and $x^-$ are
the LC time and longitudinal coordinate.

In the infinite-momentum  frame, the parton interpretation makes sense and 
deep inelastic scattering (DIS) proceeds via the
instantaneous absorbtion of the virtual photon $\gamma^*$ 
(with 4-momentum $q^\mu$) by some parton in the hadron. 
The Bjorken $x_B$ variable is defined as
$x_B\equiv Q^2/2P\cdot q$, where $Q^2 \equiv -q^\mu q_\mu$, and $P^\mu=
\delta^{\mu +}P^+$ with large $P^+$ is the hadron 
4-momentum. By kinematics, $x_B$ coincides with
the longitudinal momentum fraction $x \equiv p^+/P^+$ of the
struck parton: $x_B = x$. 

At $x\ll 1$, the gluon density increases faster,
and is the driving force towards saturation. 
The dynamics responsible for this increase is the BFKL evolution
\cite{BFKL},
as I briefly recall now 
(see also Ref. \cite{AM1} and Refs. therein):

Let me call {\em soft}, respectively {\em fast}, a parton
having small, respectively, large longitudinal momentum
(this separation will be made more precise in a moment).
A soft gluon, with $p^+=xP^+\ll P^+$, is a shortlived 
excitation which is typically radiated by a fast parton 
(e.g., a valence quark) with a larger
longitudinal momentum $k^+\gg p^+$, and thus a longer
lifetime. Indeed, by the uncertainty principle,
the lifetime of the parton system in Fig. 1.a is:
\beq
\Delta x^+\,=\,\frac{1}{\varepsilon_{k-p}+\varepsilon_p-
\varepsilon_k}\,\simeq\,\frac{1}{\varepsilon_p}\,=\,
\frac{2p^+}{p_\perp^2}\,\propto\,x,\eeq
where $\varepsilon_k\equiv k_\perp^2/2k^+$ is the LC
energy of the on-shell excitation with momentum ${\vec k}=(k^+,
{\bf k}_\perp)$, and I have used the fact that, for comparable
transverse momenta $k_\perp$ and $p_\perp$, 
$\varepsilon_p\gg\varepsilon_k,\varepsilon_{k-p}$.
To conclude, softer partons have larger energies, and
therefore shorter lifetimes.

The lowest-order process in Fig. 1.a is amended by radiative
corrections enhanced by the large rapidity gap
$\Delta \tau\equiv \ln(k^+/p^+) \sim \ln(1/x)$. 
(I assume here that $\ln(1/x)\gg  1$.)
For instance,
the probability for the emission of a second gluon with
momentum $k_1^+$ in the range $k^+ > k_1^+ > p^+$ is
(cf. Fig. 1.b) : 
\beq\label{one-gluon}
\Delta P\,\propto \,\alpha_s\int_{p^+}^{k^+}\,\frac{d k_1^+}{k_1^+}
\,=\,\alpha_s\,\ln\,\frac{k^+}{p^+}\,\sim \,\alpha_s\ln\frac{1}{x}\,,
\eeq
and becomes exceedingly large as $x\to 0$. It is then
highly probable that more gluons will be emitted along the way,
thus giving birth to the gluon cascade depicted in Fig. 1.c.
For a fixed number $N$ of gluons in this cascade,
the largest contribution, of order $(\alpha_s\ln(1/x))^N$,
comes from the kinematical domain where
\beq\label{HL}
k^+\equiv k_0^+\,\gg\,k_1^+\,\gg\,k_2^+\,\gg\,\cdots\,\gg\,
k_N^+\equiv p^+.\eeq
Other momentum orderings give contributions which are suppressed by,
at least, one factor of $1/\ln(1/x)$, and thus can be neglected
to {\em leading logarithmic accuracy} (LLA). For the dominant
contribution in eq.~(\ref{HL}), the number
of radiated gluons {\em increases exponentially} with 
$\Delta \tau\/$: $N(x)\sim {\rm exp}\{A\alpha_s
\ln(1/x)\}$, with constant $A$. This is a coherency effect,
consequence of the separation of scales in eq.~(\ref{HL}),
and of the bosonic nature of the gluons:

\begin{figure}
\protect\epsfxsize=11.5cm{\centerline{\epsfbox{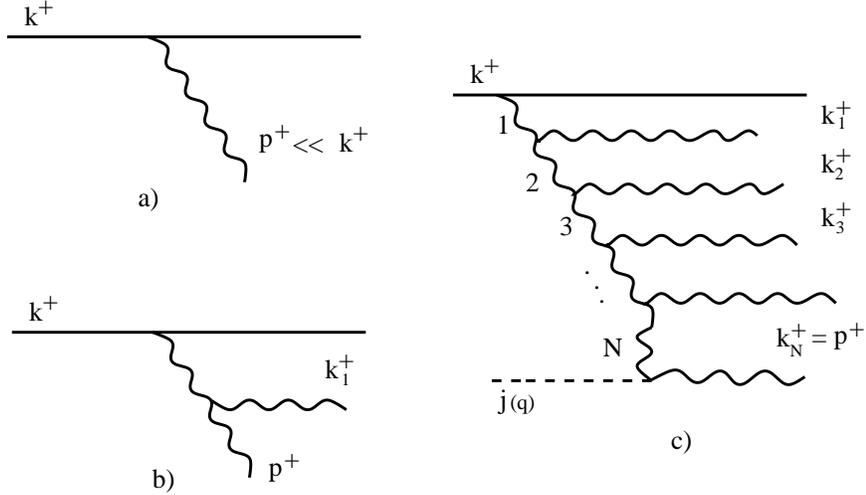}}}
         \caption{a) Soft gluon emission by a fast parton;
b) a second emission; c) a gluon cascade.}
\label{Born}
\end{figure}

Indeed, because of its short lifetime,
the soft gluon at the lower end of the cascade 
 ``sees'' the $N$ previous gluons as
a frozen color charge distribution, with an average color charge
$Q\equiv \sqrt{\langle Q_aQ_a\rangle} \propto N\/$. Thus,
the $N$th gluon is emitted {\em coherently} off the color
charge fluctuations of the previously emitted gluons, 
with a differential probability (compare to eq.~(\ref{one-gluon})) :
\beq
d P_N\,\propto \,\alpha_s\,N\,d\tau_N,\eeq
which implies that $N(\tau)\sim {\rm e}^{A\alpha_s
\tau}$, as anticipated.
Then, also the gluon density $xG(x,Q^2)\equiv (dN/d\tau) \propto
{\rm e}^{A\alpha_s\tau}$ grows exponentially\footnote{A more refined treatment,
using the BFKL equation \cite{BFKL}, gives $A=4N_c\ln 2/\pi$.}
 with $\tau\equiv \ln(1/x)$.

Thus, the BFKL picture is that of an unstable growth of the
color charge fluctuations as $x$ becomes smaller and smaller.
However, this evolution assumes the radiated gluons to behave as free 
particles, so it ceases to be valid at very low $x$, where the gluon density 
becomes so large that the radiated gluons overlap each other in the 
transverse plane and start interacting. This is the onset of saturation.

This is also the regime where the description in terms of 
a Colored Glass Condensate becomes appropriate: Because of the
hierarchy of scales in eq.~(\ref{HL}), the soft gluons
``see'' the fast partons as an effective color charge
which is {\it static} (i.e., independent
of $x^+$), and {\it localized}  near the LC (i.e.,
at $x^-=0$), with a {\it random} density $\rho_a(x^-,x_\perp)$.
(This is random since the soft gluons can belong to different
cascades, and the instantaneous configuration of the cascades
inside the hadron is random. Of course, the integrated color charge
over the whole hadron must vanish, because of confinement.)
The spatial correlations of the effective 
charge $\rho_a(x^-,x_\perp)$ reflect the quantum
dynamics of the fast partons, and are encoded in a statistical 
weight function  $W[\rho]$.

This color charge acts as a source for the soft gluons which,
because of their large occupation numbers, can be treated
in the {\it classical approximation}. This suggests the following
{\it effective theory} for the soft gluon correlators, originally
proposed by McLerran and Venugopalan \cite{MV}:
First, one solves the classical Yang-Mills equations with the 
source $\rho_a$, and in the LC-gauge\footnote{This
gauge choice will be motivated in Sect. 3.2.} $A^+=0$ :
\be
[D_{\nu}, F^{\nu \mu}]\, =\, \delta^{\mu +} \rho_a(x^-,{\bf x}_\perp)\,.
\label{cleq0}
\ee
In the saturation regime, $A\sim \rho\sim 1/g$ (see Sect. 3.2
below), and eq.~(\ref{cleq0}) must be solved {\it exactly}:
the classical problem is fully non-perturbative.
The non-linear effects 
correspond to interactions among the soft gluons, to all orders.
The corresponding solution $A^\mu\equiv {\cal A}^\mu[\rho]$
will be constructed in Sect. 3.1.
Then, the soft correlation functions of interest are obtained as:
\be\label{clascorr}
\langle A^\mu_a(x)A^\nu_b(y)
\cdots\rangle_\Lambda\,=\,
\int {\cal D}\rho\,\,W_\Lambda[\rho]\,{\cal A}_a^\mu({ x})
{\cal A}_b^\nu({ y})\cdots\,.\ee
Note the scale $\Lambda\equiv \Lambda^+$ in eq.~(\ref{clascorr}): this
is the separation scale between {\it fast} ($p^+>\Lambda^+$) 
and {\it soft} ($p^+ <\Lambda^+$) degrees of freedom.
Clearly, the structure and the correlations of $\rho$ will
depend upon $\Lambda$ (e.g., $\rho$ has support at $x^-\simle
1/\Lambda^+$, and $W[\rho]\equiv W_\Lambda[\rho]$). 
Moreover, the soft correlations must
be evaluated at longitudinal momenta $p^+$ which are not
too small as compared to $\Lambda^+$: otherwise,
the quantum corrections due to the ``semi-fast'' gluons with 
$k^+$ momenta in the range $p^+ < k^+ <  \Lambda^+$ 
would be relatively large, of O$(\alpha_s\ln(\Lambda^+/p^+))$, 
since enhanced by the large rapidity interval $\Delta\tau=
\ln(\Lambda^+/p^+)\gg 1$.

Remarkably, the equations (\ref{cleq0})--(\ref{clascorr})
are those for a glass (here,
a {\it colored} glass): There is a source, 
and the source is averaged over. 
This is entirely analogous to what is done for
spin glasses when averaging over stochastic
magnetic fields
\cite{Parisi}. By computing the two-point function as above, 
one has found \cite{JKMW97,KM98} a saturation regime where
$\langle A^i_aA^i_a\rangle_\Lambda \propto 
1/\alpha_s$, a value typical for a condensate
(see also Sect. 3.2 below). We thus
conclude that the matter which describes the small 
$x$ part of a hadron
wavefunction is a Colored Glass Condensate.

\section{Saturation in the McLerran-Venugopalan model}

In this section, I present the solution to
eq.~(\ref{cleq0}) and explore its consequences for the 
saturation of gluons.

\subsection{The non-Abelian Weizs\"acker-Williams field}

Consider first the Abelian version of eq.~(\ref{cleq0}), as a warming up:
\be
\partial _{\nu} F^{\nu \mu} =  \delta^{\mu +} \rho({\vec x}),
\label{acleq}
\ee
where $A^+=0$, $F^{\mu \nu} = \partial^\mu A^\nu-\partial^\nu A^\mu$,
and ${\vec x}\equiv (x^-,{\bf x}_{\perp})$.
Eq.~(\ref{acleq}) can be easily solved in momentum space,
with the result that ${\cal A}^-=0$ and 
\be
{\cal A}^i(p)\, =\, -{p^i \over p^+} {\rho(p^+,p_{\perp}) 
\over p_{\perp}^2}\,.
\label{aaimom}
\ee
Here, one needs a prescription to handle the pole at $p^+=0\/$:
different prescriptions 
correspond to different boundary conditions in $x^-$.
For instance, with the 
``retarded'' prescription\footnote{Note that the
``retardation'' refers here to $x^-$, and not to the LC 
time $x^+$.} $1/(p^+\ +i \varepsilon)$, one obtains 
 \be
{\cal A}^i (x^-,x_{\perp})=
\int_{-\infty}^{x^-} dy^- \, \partial^i \alpha (y^-,x_{\perp})\,,
\label{aiab1}
\ee
with $\alpha$ satisfying
$-\nabla_{\perp}^2 \alpha ({\vec x}) = \rho ({\vec x})$.

The vector potential (\ref{aiab1}) is static, $\partial^- {\cal A}^i\equiv
(\partial {\cal A}^i/\partial x^+) =0$, and defines a two-dimensional 
pure gauge: ${\cal F}^{ij}=0$. But, of course, this is not
a pure gauge in {\em four} dimensions, since the electric field 
${\cal F}^{i+}=-\partial^+ {\cal A}^i =\partial^i\alpha\,$ is
non-zero. The above solution, also
known as the Weizs\"acker-Williams field, is the analog
of the Coulomb field in the infinite momentum frame.

It turns out that
the corresponding non-Abelian solution has a similar structure:
Specifically, the solution ${\cal A}^\mu$ to eq.~(\ref{cleq0}) 
in the LC gauge (${\cal A}^+=0$) can be chosen
such as ${\cal A}^-=0$, and ${\cal A}^i$ is static and
a two-dimensional pure gauge: ${\cal F}^{ij}=0$. This
can be written as follows
(for retarded boundary conditions):
\be\label{tpg} 
{\cal A}^i(x^-,x_{\perp})&=&(i/g)
U(x^-,x_{\perp})\,\partial^i U^{\dagger}(x^-,x_{\perp})\\
U^{\dagger}(x^-,x_{\perp})&=&
 {\rm P} \exp
 \left \{
ig \int_{-\infty}^{x^-} dz^-\,\alpha (z^-,x_{\perp})
 \right \},\label{UTA}\\
\label{EQTA}
- \nabla^2_\perp \alpha({\vec x})&=&{\tilde \rho}(\vec x)\,
\equiv\,U^{\dagger}(\vec x)
\, \rho(\vec x) \,U(\vec x),\ee
where P is the usual path-ordering operator,
and ${\tilde \rho}(\vec x)$ is the value of the
classical source in the {\it covariant gauge}
 $\partial_\mu {\tilde A}^\mu =0$.

Note that, while $\alpha$ is linearly related to ${\tilde \rho}$,
its relation to the LC-gauge source $\rho$ is more 
complicated, since the gauge rotations $U$ and $U^{\dagger}$
are themselves functionals of
$\alpha$, cf. eq.~(\ref{UTA}). Thus, in order to perform the
average in eq.~(\ref{clascorr}), 
it is more convenient to use ${\tilde \rho}$ (rather than $\rho$) 
as the independent variable. This is possible because the measure 
and the weight function in eq.~(\ref{clascorr})
are gauge invariant; e.g., $W_\Lambda[\rho]=W_\Lambda[\tilde\rho]$.
Then, the non-Abelian solution
${\cal A}^i[\tilde \rho]$ is known
explicitly, via eqs.~(\ref{tpg})--(\ref{EQTA}), and
eq.~(\ref{clascorr}) is replaced by
\be\label{COVclascorr}
\langle A^i(x^+,\vec x)A^j(x^+,\vec y)
\cdots\rangle_\Lambda\,\,=\,
\int {\cal D}\tilde\rho\,\,W_\Lambda[\tilde\rho]\,\,{\cal A}^i_{\vec x}
[\tilde \rho]\,{\cal A}^j_{\vec y}[\tilde \rho]\cdots\,.
\ee
In particular, this equation shows that only the
{\it equal-time} ($x^+=y^+$)
correlators of the soft {\it transverse} ($\mu=i$, $i=1,2$)
fields can be computed in this model; but these are precisely
the gluon correlators 
which are probed in DIS. 

Recall finally that the source $\tilde\rho$ is created by the fast
partons with $|p^+|>\Lambda^+$, so its support is restricted
to $|x^-|\simle 1/\Lambda^+$.
The small-$x$ external probe, on the other hand,
is sensitive only to the gross
features of the color fields ${\cal A}^i$
over large distances $|x^-| \gg 1/\Lambda^+$, where
one can replace eq.~(\ref{UTA}) with
\be\label{UTAF}
U^{\dagger}(x^-,x_{\perp})\,\approx\,\theta(x^-)
 {\rm P} \exp
 \left \{
ig \int_{-\infty}^{\infty} dz^-\,\alpha (z^-,x_{\perp})
 \right \}\,\equiv \,\theta(x^-)\,v^\dagger(x_{\perp}),
\ee
and therefore (cf. eq.~(\ref{tpg})):
\be\label{APM}
{\cal A}^i\approx\theta(x^-)\,
\frac{i}{g}\,v(\del^i v^\dagger)
\,\equiv\,\theta(x^-){\cal A}^i_\infty(x_\perp),\qquad
{\cal F}^{i+}\approx-\delta(x^-)\,
{\cal A}^{i}_\infty(x_\perp),\ee
where, strictly speaking, the $\delta$-function 
is localized at $ x^- \simle 1/\Lambda^+$.

\subsection{Gluon distribution function and saturation}

Consider the simple approximation where the 
weight function for $\tilde\rho$, which is not yet known,
is taken to be a Gaussian \cite{MV} :
\be\label{FCLAS}
W_\Lambda[\tilde\rho]\simeq \exp\left\{-
\frac{1}{2}\int d^3 x \,\frac{\tilde\rho_a^2(\vec x)}
{\xi^2_\Lambda(\vec x)}\right\}\,,\ee
where $\xi^2_\Lambda$ is  the total color charge 
density squared of the partons with $p^+>\Lambda^+$.
By using this approximation, it has been possible to compute 
the gluon distribution function $xG(x,Q^2)$ in the MV model
\cite{JKMW97,KM98}, as I recall now:

By definition, $G(x,Q^2)$ is the number of gluons
in the hadron wavefunction having 
longitudinal momentum $k^+=xP^+$, and transverse momentum less than $Q$.
In the LC gauge, this is simply related to the
Fock-space gluon distribution function\footnote{Of course, the 
Fock-space gluon distribution can be defined in any gauge;
but it is only in the LC-gauge that the definition (\ref{GDF0}) can be
given a gauge invariant meaning \cite{AM1,CGC}.}, and therefore to the
gluon two-point function \cite{AM1}:
\be\label{GDF0}
G(x,Q^2)&\equiv&
\int {d^2k_\perp \over (2 \pi)^2}\,\Theta(Q^2-
k_\perp^2)\int{dk^+ \over 2 \pi}
\,2k^+\,\delta\biggl(x-{k^+\over P^+}\biggr)\nonumber\\
&{}&\qquad\qquad
\times\,\Bigl\langle A^i_a(x^+,k^+,{\bf k}_\perp)
A^i_a(x^+,-k^+,-{\bf k}_\perp)\Bigr\rangle,\,\,\,\ee
where the brackets denote the average over the hadron wavefunction.
In the LC-gauge gauge, $F^{i+}(k)=ik^+A^i(k)$, and therefore
(with $k^+=xP^+$) :
\be\label{GDF}
xG(x,Q^2)\,=\,\frac{1}{\pi}\int {d^2k_\perp \over (2 \pi)^2}\,
\Theta(Q^2-k_\perp^2)\,\Bigl\langle F^{i+}_a(x^+,\vec k)
F^{i+}_a(x^+,-\vec k)\Bigr\rangle.\ee
If the integration over $k_\perp$ were unrestricted,
this quantity would be manifestly gauge-invariant.
But even for a finite $Q^2$, this
has a gauge-invariant meaning when evaluated on the
non-Abelian Weizs\"acker-Williams field in Sect. 3.1
 \cite{CGC}. In this approximation,
$F^{i+}(x^+,\tilde k)\approx {\cal F}^{i+}(k_\perp)
=- {\cal A}^{i}_\infty(k_\perp)$ (cf. eq.~(\ref{APM})), and 
\be
\label{GCL}
x G(x,Q^2)&\approx&
{R^2}\int^{Q^2} {d^2k_\perp \over (2 \pi)^2}
\int d^2x_\perp\,{\rm e}^{-ik_\perp\cdot x_\perp}
\Bigl\langle  {\cal A}^{ia}_\infty(0)\,
 {\cal A}^{ia}_\infty(x_\perp)\Bigr\rangle_\Lambda,\ee
where $R$ is the hadron radius (I have assumed
homogeneity in the transverse plane, for simplicity),
and the average is to be understood 
in the sense of eq.~(\ref{COVclascorr}) with $\Lambda^+=xP^+$.
Thus, the r.h.s. of eq.~(\ref{GCL}) is still dependent
on $x$, but only via the respective dependence of the
weight function for $\tilde\rho$. 

With the Gaussian weight function (\ref{FCLAS}), and the non-linear
classical solution in Sect. 3.1, the gluon distribution
(\ref{GCL}) can be computed exactly \cite{JKMW97,KM98}:
\be\label{SATN}
\Bigl\langle  {\cal A}^{ia}_\infty(0)\,
 {\cal A}^{ia}_\infty(x_\perp)\Bigr\rangle
\,=\,\frac{N_c^2-1}{\pi\alpha_s N_c}\,\frac{1-{\rm e}^{-x_\perp^2
\ln(x_\perp^2 \Lambda_{QCD}^2) Q_s^2/4}}
{x_\perp^2}\,,\ee
where $N_c$ is the number of colors, and
$Q_s\propto \alpha_s\xi_\Lambda$ is the {\it saturation momentum}
and is a function of $\Lambda^+$, that is, of Bjorken's $x$. 
This  equation displays
saturation: the vector potential never becomes larger than
$A^i\sim 1/g$. This is the maximal occupation number permitted
for a classical field, since larger occupation numbers are blocked by 
repulsive interactions of the gluon field.

This interpretation can be made sharper by going to momentum space:
If $N(k_\perp)$ is the Fourier transform of (\ref{SATN})
[this is the same as $(dN/d^2k_\perp\,d\tau\,d^2x_\perp)$,
the gluon density per unit rapidity and unit
 transverse phase-space], then:
\be N(k_\perp) \,\propto\, \alpha_s (Q_s^2/k_\perp^2) \qquad{\rm for}
\quad k_\perp^2\gg Q_s^2,\ee
which is the normal perturbative behavior, but\footnote{In relativistic
heavy ion collisions, one expects \cite{AM2} $Q_s\sim 1$ GeV
at RHIC, and $Q_s\sim 2-3$ GeV at LHC. Thus, at least at LHC,
the kinematical window in eq.~(\ref{NSAT}) should be non-negligible.}
\be\label{NSAT}
 N(k_\perp) \,\propto\, {1\over \alpha_s}\,\ln\,
\frac{k_\perp^2}{Q_s^2}\qquad{\rm for}\quad 
\Lambda_{QCD}^2\ll k_\perp^2\ll Q_s^2,\ee
which shows a much slower increase, i.e., saturation, at
low momenta. 

Note, however, that the above argument is not rigurous, since
the local Gaussian form for $W[\rho]$ in eq.~(\ref{FCLAS}) is 
valid only at
sufficiently large transverse momentum scales so that the effects of
high gluon density are small. It is therefore important to verify
if saturation comes up similarly with a more realistic form for
the weight function, as obtained after including the
quantum evolution in $x$. This will be discussed now.

\section{The non-linear evolution equation}

Because the separation of scales is only logarithmic,
the effective theory (\ref{cleq0})--(\ref{clascorr})
with scale $\Lambda^+$ applies only to gluon correlations
at a scale $p^+$ slightly below $\Lambda^+$. 
If one is interested in correlations at the
softer scale $b\Lambda^+$ with $b\ll 1$, then, to LLA,
one has to include also the corrections of order 
$\alpha_s\ln(1/b)$ due to the {\it semi-fast} quantum 
fluctuations with longitudinal momenta in the strip
\be\label{strip}\,\,
 b\Lambda^+ \,\,<\,\, |k^+|\,\, <\,\,\Lambda^+\,.\ee
Together with the dynamical information already
contained in the effective theory at scale $\Lambda^+$
(``Theory I''), these additional corrections will determine 
the effective theory at the softer scale $b\Lambda^+$ 
(``Theory II''). 
This suggests an iterative construction of the effective 
theory where the quantum fluctuations are integrated out
in layers of $k^+$, down to the physical scale of interest.
At each step in this procedure, the quantum corrections must be
computed to leading order in $\alpha_s\ln(1/b)$ (LLA),
but to {\em all} orders in the strong background field
${\cal A}^i\sim 1/g$ 
produced by the color source $\rho$ at the previous step.

To compute quantum corrections, one needs the quantum
generalization of the McLerran-Venugopalan model \cite{JKLW1,JKLW2,JKW}. 
This is obtained by replacing eq.~(\ref{clascorr}) with 
(below, the gauge condition $A^+=0$ is implicit):
\be\label{2point}
\langle {\rm T}
A^\mu(x)A^\nu(y)\rangle
\,=\,\int {\cal D}\rho\,\,W_\Lambda[\rho]
\left\{\frac{\int^\Lambda {\cal D}A
\,\,A^\mu(x)A^\nu(y)\,\,{\rm e}^{\,iS[A,\,\rho]}}
{\int^\Lambda {\cal D}A\,\,{\rm e}^{\,iS[A,\,\rho]}}
\right\},\ee
which involves {\em two} functional integrals: a quantum 
path integral over the soft ($k^+< \Lambda^+$) 
gluon fields $A^\mu$, which defines quantum expectation
values at fixed $\rho$, and a classical average 
over $\rho$, with weight function $W_\Lambda[\rho]$.
(This double averaging is similar to the one
performed for spin systems at finite temperature and
in a random external magnetic field \cite{Parisi}.)
Unlike eq.~(\ref{clascorr}), the 2-point function
given by eq.~(\ref{2point}) is independent of the arbitrary separation
scale $\Lambda^+\/$: the cutoff dependence of the quantum loops cancels 
against the corresponding dependence of the classical weight function
$W_\Lambda[\rho]$.

The action $S[A,\rho]$ is chosen such as to be gauge-invariant and 
reproduce the classical equations of motion (\ref{cleq0})
in the saddle point approximation $\delta S/\delta A^\mu=0$.
It reads \cite{JKLW1}: $S=S_{YM}+S_W$, where
$S_{YM}=\int d^4x(-F_{\mu\nu}^2/4)$ is the Yang-Mills action, and
$S_W$ is a gauge-invariant generalization of the
eikonal vertex $\int d^4 x \rho A^-$ 
 (with T denoting time-ordering of color matrices) :
\be\label{SW}
S_W\,\equiv\,
{i \over {gN_c}} \int d^3 \vec x\,\, {\rm Tr}\,\left\{ \rho(\vec x)
\,{\rm T}\, \exp\left(\,
ig\int dx^+ A^-(x^+,\vec x) \right)\right\}\,.\ee

\begin{figure}
\protect\epsfxsize=8.5cm{\centerline{\epsfbox{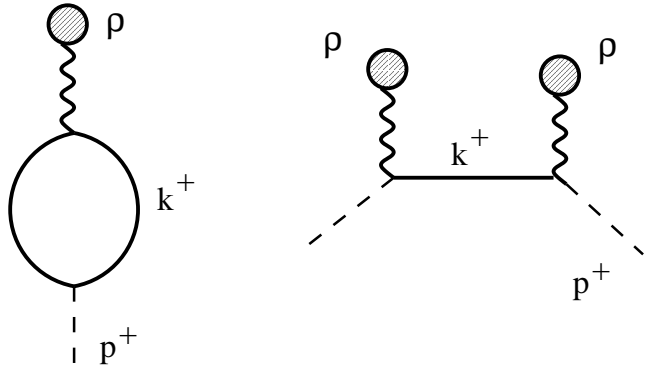}}}
         \caption{Contributions to $\sigma$ (a)
and $\chi$ (b) to lowest order in $\rho$. Wavy lines with a blob
denote insertions of $\rho$; the continuous lines are propagators
of the semi-fast gluons; the external, dotted, lines
carry soft momenta.}
\label{LINSC}
\end{figure}

Theories I and II are defined as in eq.~(\ref{2point}), but with
separation scales equal to $\Lambda^+$, and $b\Lambda^+$,
respectively. The difference $\Delta W
\equiv W_{b\Lambda} - W_\Lambda$ can be obtained by
matching calculations of gluon correlations at the scale 
$p^+\simle b\Lambda^+$ in the two theories \cite{CGC}.
In Theory II, and to lowest order in $\alpha_s$,
these correlations are found at tree level, 
i.e., in the classical approximation.
In Theory I, and to the same accuracy,
they involve also the logarithmically enhanced quantum
corrections due to the ``semi-fast'' fluctuations defined
by eq.~(\ref{strip}). The result can be
expressed as an evolution equation\footnote{This equation
insures that correlation functions like 
(\ref{2point}) are independent of  $\Lambda^+$.}
for $W_\tau[\rho]$
(with $\tau\equiv \ln(P^+/\Lambda^+)$) with respect to variations
in $\tau$, originally derived in Ref. \cite{JKLW2}. This reads:
\be\label{RGE}
{\del W_\tau[\rho] \over {\del \tau}}\,=\,
 \alpha_s \left\{ {1 \over 2} {\delta^2 \over {\delta
\rho_x \delta \rho_y}} [W_\tau\chi_{xy}] - {\delta \over {\delta \rho_x}}
[W_\tau\sigma_x] \right\}\,.
\ee
in condensed notations where (with
$x^-_\tau\equiv 1/\Lambda^+ = {\rm e}^{\tau}x_0^-$), e.g.,
\be\label{rhoW3}
\frac{\delta}{\delta \rho_x}\,[W_\tau\sigma_{x}] \,=\,
\int d^2{x_\perp} \,\frac{\delta}{\delta \rho_a(x^-_\tau,x_\perp)}\,
\Bigl[W_\tau\sigma_a(x_\perp)\Bigr],\ee
 and the 
functions $\sigma_a(x_\perp)$ and $\chi_{ab}(x_\perp,y_\perp)$
are the one- and two-point correlators of the color charge
$\delta\rho_a(x)$ of the semi-fast  quantum fluctuations:
\be\label{SIGCHI}
\alpha_s\ln{1\over b}\,\sigma_a ({ x}_\perp)&\equiv&
\int dx^- \,\langle\delta \rho_a (x^+,x^-,{ x}_\perp)
\rangle_\rho,\\
\alpha_s\ln{1\over b}\,
\chi_{ab}({ x}_\perp,{ y}_\perp)&\equiv&
 \int dx^- \int dy^-\,\langle \delta \rho_a (x^+,x^-,{ x}_\perp) 
\delta \rho_b (x^+,y^-,{y}_\perp)\rangle_\rho\,.\nonumber\ee
Thus, $\sigma$ and $\chi$ are (generally non-linear) functionals of 
$\rho$ obtained by averaging over the semi-fast quantum
fluctuations in the presence of the tree-level fields and sources,
${\cal A}^i$ and $\rho$. To the order of interest,
the semi-fast fields can be integrated out in the Gaussian
approximation; thus, the Feynman graphs contributing to
$\sigma$ and $\chi$ involve, at most, one loop.
Some typical contributions to lowest order in $\rho$
are shown in Fig. \ref{LINSC}. As one can see on these figures,
$\sigma$ is a one-loop virtual correction to the tree-level source
and describes vacuum polarisation, while $\chi$ is a tree
diagram which describes the real emission of a
semi-fast gluon (cf. Fig 1.b).

The functional equation (\ref{RGE})
is equivalent to an infinite hierarchy of 
ordinary equations for the correlators of the charge density.
For instance, by multiplying eq.~(\ref{RGE}) with $\rho_x\rho_y$
and functionally integrating over $\rho$, one obtains an evolution
equation for the two-point function:
\be\label{RGE2p}
{d\over {d\tau}}\,
\langle\rho_x\rho_y\rangle_\tau\,=\, \alpha_s\,\langle\chi_{xy}
+\rho_x\sigma_y+\sigma_x\rho_y\rangle_\tau\,,\ee
which in general involves also the higher $n$-point 
functions, via $\sigma$ and $\chi$.
But in the limit of a weak source, i.e., with $\sigma$ and $\chi$
computed to lowest order in $\rho$, this becomes a closed 
equation for $\langle\rho\rho\rangle$ which coincides with
the BFKL equation\cite{JKLW1}, as necessary on physical grounds.
This is a very non-trivial check of the effective theory
in eqs.~(\ref{2point})--(\ref{SW}).
Specifically, the linear (in $\rho$)
contribution to $\sigma$ depicted in Fig. 2.a
is evaluated as:
\begin{equation}\label{sig1x}
\langle\delta \rho_a (\vec x)
\rangle_\rho \, = \,
{g^2 N_c\over 2 \pi} \,F(x^-)
\int \frac{d^2 p_\perp}{(2\pi)^2}
\int \frac{d^2 k_\perp}{(2\pi)^2}\,{\rm e}^{i(p_\perp-k_\perp)
\cdot x_\perp}\,
{p_{\perp} \cdot k_{\perp} \over p_{\perp}^2 k_{\perp}^2}
\,\rho^a(q_\perp),
\end{equation}   
with the ``form factor''
\be\label{FormF}
F(x^-)\,\equiv\,\theta(x^-)\,\frac{e^{-ib\Lambda^+ x^-}-
e^{-i\Lambda^+ x^-}}{x}\,,\ee
which has support at 
$1/\Lambda^+ \simle x^- \simle 1/b\Lambda^+\/$, and yields
the expected logarithmic enhancement (cf. eq.~(\ref{SIGCHI}))
after the integration over $x^-\/$:
\be 
\int dx^- F(x^-) \,=\,\ln {1 \over b}\,.\ee
The kernel of the two-dimensional integral
in eq.~(\ref{sig1x}) can be recognized as
the {\em virtual} part of the BFKL kernel; the corresponding 
{\em real} part is generated by $\chi$ when evaluated to lowest order
in $\rho$ (cf. Fig. 2.b) \cite{JKLW1}.

The general, non-linear, expressions for $\sigma$ and $\chi$
valid in the saturation regime
will be presented somewhere else \cite{CGC}.
(See also Ref. \cite{JKW} for an alternative calculation,
with different results though.) 
Here, I would like only to emphasize the longitudinal
structure of the quantum corrections, which is
already manifest on the linear approximation 
(\ref{sig1x})--(\ref{FormF}) : Since generated by modes
with $k^+$ momenta in the strip (\ref{strip}), the induced source
$\langle\delta \rho\rangle$
is localized at $1/\Lambda^+ \simle x^- \simle 1/b\Lambda^+$, 
that is, on top of the tree-level source $\rho$ (which has support
at $0 \simle x^- \simle 1/\Lambda^+$).
Because of that, the functional derivatives in 
eqs.~(\ref{RGE})--(\ref{rhoW3}) are to
be evaluated at $x^-=x^-_\tau\equiv 1/\Lambda^+$ : the
evolution from $W_\tau[\rho]$ to $W_{\tau+d\tau}[\rho]$
is due to changes in $\rho$ within the rapidity interval
$(\tau, \tau+d\tau)$.

In fact, the precise longitudinal picture depends upon the
$i\epsilon$ prescription for the ``axial'' pole $1/p^+$ 
in the LC-gauge propagator of the semi-fast 
gluons\footnote{This prescription is necessary for fixing the
residual gauge freedom in the LC-gauge.}.
The result in eq.~(\ref{FormF}) has been obtained \cite{CGC}
by using the retarded prescription $1/(p^+\ +i \varepsilon)$, 
for consistency with the classical solution
(\ref{tpg})--(\ref{EQTA}).
If the advanced prescription $1/(p^+\ -i \varepsilon)$ was
used instead, the corresponding
$\langle\delta \rho_a (\vec x)\rangle$
would be located at negative $x^-$, but its integral over $x^-$
--- which defines $\sigma(x_\perp)$, cf. eq.~(\ref{SIGCHI})
--- would nevertheless be the same. That is, the BFKL equation
is obtained independently of the axial prescription.

This is, however, specific to the linearized (or BFKL) 
limit of the evolution equation.
In general, the non-linear effects depend upon 
the gauge condition, as shown by the explicit calculations
of $\sigma$ and $\chi$ using different prescriptions \cite{JKW,CGC}.
The simplest results \cite{CGC} are obtained by using the retarded,
or the advanced, prescriptions alluded to before. With
a {retarded} prescription, both the classical mean field
${\cal A}^i_a$, eq.~(\ref{APM}), and the induced source
$\langle\delta \rho_a \rangle$, eq.~(\ref{sig1x}),
sit at $x^- > 0$, or $z < t\,$; that is, the soft gluons fields
are behind their source, the hadron (which is located at $z=t$). 
It is intuitively plausible
that within this picture there are no initial state gluon interactions.
This is the counterpart of the conclusion of Mueller and
Kovchegov \cite{KM98,AM2} who used an {advanced} prescription 
$1/(p^+-i\epsilon)$, and showed that all the {\em final}
state gluon interactions disappeared.

These two prescriptions are simple in that one
can either put gluon interactions in the final or initial state. 
Other gauge conditions such as
Leibbrandt-Mandelstam or principle value do not have this simple feature,
and lead to other complications for computations as well \cite{CGC}. 
Of course, gauge invariant quantities must come out the same whatever
gauge is used for their computation.
But the classical color charge $\rho_a(\vec x)$ is not a
physical observable by itself, but just a convenient tool to 
summarize the (generally, gauge-variant) correlations 
inherited by the soft gluons from the fast partons.
The discrepancies between the results for $\sigma$ and $\chi$ 
found in Refs. \cite{JKW,KM,KMW} and Ref. \cite{CGC} may be
ultimately attributed to the different methods used to fix
the gauge. In fact, our results \cite{CGC} are
consistent with some previous results by Balitskii \cite{B}
and Kovchegov \cite{K}, to which they reduce in the large $N_c$
limit.

To conclude, the  non-linear evolution equation (\ref{RGE})
together with the expressions for $\sigma$ and $\chi$ to be
presented in Ref. \cite{CGC} provide a theoretical framework
to perform calculations in the saturation regime.
Further studies within this framework 
should lead to a quantitative understanding of unitarity 
in high energy scattering, and of the structure of final
states and multiparticle production \cite{MV,K}. 
This would have important applications 
to the theory of deep inelastic scattering, hadronic
interactions and ultrarelativistic nuclear collisions.

\end{document}